\documentclass[12pt,draftcls,onecolumn]{IEEEtran}
\usepackage[update,prepend]{epstopdf}
\usepackage{amsmath,rotating}
\usepackage{hyperref}
\usepackage{amsfonts}
\usepackage{amssymb}
\usepackage{times,subfigure}
\usepackage{comment}
\usepackage{epstopdf}
\usepackage{algorithmic}
\usepackage{algorithm}
\usepackage{graphicx}
\usepackage[T1]{fontenc}
\usepackage[scanall]{psfrag}
\newtheorem{Theorem}{Theorem}
\newtheorem{Corollary}{Corollary}

\newtheorem{Remark}{Remark}

\def\sqw{\hfill\hbox{\lower.1ex\hbox{$\sqcup$}
    \kern-1.02em\lower.1ex\hbox{$\sqcap$}}\ }

\DeclareMathOperator*{\argmin}{\arg\min}
\DeclareMathOperator*{\argmax}{\arg\max}

\DeclareMathOperator*{\diag}{diag}

\newcommand{\qed}{\hfill \mbox{\raggedright \rule{.07in}{.1in}}}

\newcommand{\yv}{\mathbf{y}}
\newcommand{\av}{\mathbf{a}}
\newcommand{\bv}{\mathbf{b}}

\newcommand{\xv}{\mathbf{x}}

\newcommand{\pv}{\mathbf{p}}

\newcommand{\qv}{\mathbf{q}}

\newcommand{\RM}{\mathbf{R}}

\newcommand{\R}{\mathbb{R}}

\title{On the Randomized Kaczmarz Algorithm}

\author{
\IEEEauthorblockN{
Liang Dai,
Mojtaba Soltanalian,
Kristiaan Pelckmans\\
\IEEEauthorblockA{Department of Information Technology, Uppsala University, Sweden.}
\thanks{Research in this paper was supported by the VR project with number 106400801, and the European Research Council (ERC) grant with number 228044. }}}
\date{}

\begin{document}

\maketitle
\begin{abstract}
    The Randomized Kaczmarz Algorithm is a randomized method which aims at solving a consistent system of over determined linear equations. This note discusses how to find an optimized randomization scheme for this algorithm, which is related to the question raised by \cite{c2}. Illustrative experiments are conducted to support the findings.
\end{abstract}


\IEEEpeerreviewmaketitle

\begin{section}{Problem Statement}
In this note, we discuss the Kaczmarz Algorithm (KA)\cite{c4}, in particular the Randomized Kaczmarz Algorithm (RKA) \cite{c1}, to find the unknown vector $\xv\in \mathbb{R}^{n}$ of the following set of {\em consistent } linear equations:
\begin{align}
A\xv = \bv,
\label{eq1}
\end{align}
where matrix $A \in \mathbb{R}^{m\times n}, m\ge n$, is of full column rank, and $\bv \in \mathbb{R}^{m}$.
Since \cite{c4}, the KA has been applied to different fields and many new developments are reported. For instance, in \cite{c13}, the author study the RKA  when applied to the case of the linear systems are inconsistent. In \cite{c11}, RKA is applied to the Computer Tomography. In \cite{c14}, the authors present a method to accelerate the convergence of the RKA with the application of the Johnson-Lindenstrauss Lemma. In \cite{c15}, the authors analyze the almost sure convergence of the RKA when proper stochastic properties of matrix $A$ are introduced. In \cite{c18}, the authors presented a practically more efficient approach to solve the linear systems by projecting to different blocks of rows of $A$, and a randomization technique is applied to find a good partition of the rows.

The KA can be described as follows. Let us define the hyperplane $H_i$ as:
\begin{align*}
H_i = \{\xv| \av_i^{T}\xv = b_i\},
\end{align*}
where the $i$-th row of $A$ is denoted as $\av_i^{T}$ and the $i$-th element of $\bv$ is denoted as $b_i$. Geometrically, the solution of (\ref{eq1}) can be thought as the intersection of all hyperplanes $\{H_i\}_{i=1}^{m}$, and the KA seeks to find the solution by successively projecting to the hyperplanes from an initial approximation $\xv_0$. The process is mathematically written as
\begin{align}
\xv_{k+1} = \xv_k + \frac{b_i-\av_i^{T}\xv_k}{\|\av_i\|_2^2}\av_i,
\label{eq.rec}
\end{align}
where $i = mod(k,m) +1.$ Here we use the Matlab convention $mod(\cdot,\cdot)$ to denote the {\em modulus after the division} operation.
Fig. \ref{fig.rka} illustrates the algorithm in a low dimensional case.
\begin{figure}[htbp] %  figure placement: here, top, bottom, or page
   \centering
   \includegraphics[width=2.3in]{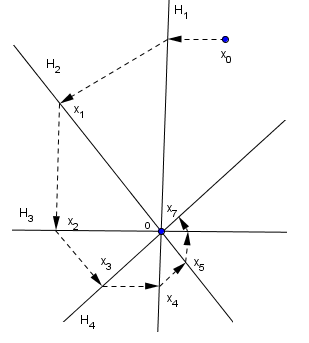}
   \caption{A geometrical interpretation of the algorithm. Here, $m=4$ and $n=2$, and the solution $\xv$ to $A\xv = \bv$ is represented by the point $o$. We can see that by this sequence of projections, $\xv_k$ converges to the solution.}
   \label{fig.rka}
\end{figure}

The key difference between the RKA and the KA is that RKA chooses the rows following a specified probability distribution. More precisely, the probability for selecting $\av_i^{T}$ is given as $\frac{\|\av_i\|_2^2}{\|A\|_F^2}.$ Note that this probability is proportional to the row norms.

 Although the KA is simple to state, its rate of convergence is still not completely explored. While for the RKA, with the predescribed  choice of the probability distribution, the following convergence result is set up in \cite{c1}:
\begin{align}
\mathbb{E}(\|\xv_k-\xv\|_2^2) \le (1-\kappa(A)^{-2})^{k}\|\xv_0-\xv\|_2^2,
\end{align}
in which $\kappa(A)= \|A\|_F\|A^{\dagger}\|_2,$ and with $\mathbb{E}$ concerning the random choices of rows in the RKA.

However, it is argued in \cite{c2} that {\em 'Assigning probabilities corresponding to the row norms is in general certainly not optimal'}.  In the follows, we will try to find an optimized probability distribution for selecting the rows from $A$, so that a better performance can be obtained. The distribution vector is derived by minimizing an upper bound to the convergence rate which can be obtained by solving a convex optimization problem.

  This note is organized as follows. The next section discusses the main results; In section 3, we discuss how to approximately solve the arising Semi-Definite-Programming (SDP) problem with smaller computational cost; In section 4, illustrative experiments will be conducted to verify the findings; Finally, we draw some conclusions in section 5.
\end{section}
\begin{section}{Optimized RKA}
In the following, for convenience of discussion, we will introduce a new matrix $B\in \R^{m\times n}.$ Let $\bv_i^{T}$ denote the $i$-th row of $B$, which is defined as

\begin{align}
\bv_i = \frac{\av_i}{\|\av_i\|_2}, \forall i = 1,\cdots,m,
\end{align}
i.e. every row of the matrix $B$ is a normalized version of the corresponding row of matrix $A$.

Let $\pv \in \mathbb{R}^{m}$ be a probability distribution vector (i.e. $\pv\ge 0$, $\mathbf{1}^{T}\pv =1$) for selecting the rows in the RKA method and let $p_i$ denote the $i$th element of $\pv$.

Assume that currently we have $\xv_{k-1}$, and based on $\xv_{k-1}$, the next approximation $\xv_{k}$ is given by (\ref{eq.rec}), in which the index $i$ is chosen randomly according to $\pv$. By the property of the projection operation, we have that
\begin{align}
\|\xv_k - \xv\|_2^2 = \|\xv_{k-1} - \xv\|_2^2\sin^2(\alpha_i),
\end{align}
in which $\alpha_i$ denotes the angle between $\xv_{k-1} - \xv$ and the selected $\bv_i$, i.e. the normal direction of the chosen hyperplane.

Based on the previous formula, we have that
\begin{align}
\mathbb{E}_{\cdot|\xv_{k-1}}(\|\xv_k - \xv\|_2^2) = \|\xv_{k-1} - \xv\|_2^2\sum_{i=1}^{m}p_i \sin^2(\alpha_i),
\label{eq.expec}
\end{align}
in which $\mathbb{E}_{\cdot|\xv_{k-1}}$ denotes the expectation operator conditioned on $\xv_{k-1}$. It follows that:
\begin{align}
\sum_{i=1}^{m}p_i \sin^2(\alpha_i)\le \sup_{\yv\in \mathbb{R}^{n},\yv\ne \mathbf{0}} \sum_{i=1}^{m}p_i \sin^2(\beta_i) \triangleq \Omega_1,
\label{eq.omega1}
\end{align}
and
\begin{align}
\sum_{i=1}^{m}p_i \sin^2(\alpha_i)\ge \inf_{\yv\in \mathbb{R}^{n},\yv\ne \mathbf{0}} \sum_{i=1}^{m}p_i \sin^2(\beta_i) \triangleq \Omega_2,
\label{eq.omega2}
\end{align}
in which $\beta_i$ denotes the angle between $\yv$ and $\bv_i$.

Based on the relations in (\ref{eq.expec}), (\ref{eq.omega1}) and (\ref{eq.omega2}), we have that
\begin{align}
\mathbb{E}_{\cdot|\xv_{k-1}}(\|\xv_k - \xv\|_2^2) \le \Omega_1\|\xv_{k-1} - \xv\|_2^2,
\label{eq.rec1}
\end{align}
and
\begin{align}
\mathbb{E}_{\cdot|\xv_{k-1}}(\|\xv_k - \xv\|_2^2) \ge \Omega_2\|\xv_{k-1} - \xv\|_2^2.
\label{eq.rec2}
\end{align}

By iterating the relations given in eq. (\ref{eq.rec1}) and eq. (\ref{eq.rec2}), the following results follow.
\begin{Theorem}
We have that
\begin{align}
\mathbb{E}(\|\xv_k - \xv\|_2^2) \le  \Omega_1^k\|\xv_0 - \xv\|_2^2,
\label{eq.bound1}
\end{align}
and
\begin{align}
\mathbb{E}(\|\xv_k - \xv\|_2^2) \ge  \Omega_2^k\|\xv_0 - \xv\|_2^2,
\label{eq.bound2}
\end{align}
in which the expectations are taken with respect to all the random choices of the rows up to time $k$.
\end{Theorem}

\begin{Remark}
Note that $\Omega_1<1$ can be guaranteed if $\pv$ is a strictly positive vector. This can be proven by a contradiction argument as follows. If $\Omega_1=1$, and since $\sin^2(\beta_i)\le 1$ for any $i$ and $\sum_{i=1}^{m}p_i=1$, we have that $\sin^2(\beta_i) = 1$, i.e. $\cos(\beta_i) = 0$ holds for all $i$. Considering that $rank(A) = n$, i.e. $rank(B) = n$, hence $\xv_k - \xv$ can not be orthogonal to the vectors $\{\bv_i\}_{i=1}^m$, and the result follows. Based on this observation, we can see that exponential convergence in expectation can be obtained by a wide range of probability distribution vectors. This finding extends the result in \cite{c1}, which only guarantees the exponential convergence for a given specific choice of the probability distribution vector. $\blacksquare$
\end{Remark}

 According to Theorem 1, in order to get a better performance, we need to find a probability distribution vector, such that $\Omega_1$ can be made as small as possible. When the optimized $\Omega_1$ is obtained, we can also have a lower bound to the convergence speed of the RKA based on $\Omega_2$. In the following, we will first derive a closed form for $\Omega_1$ and $\Omega_2$, and then introduce a convex optimization problem to calculate the probability distribution vector $\hat{\pv}$ which minimizes $\Omega_1$.

Notice that
\begin{align*}
\sum_{i=1}^{m}p_i \sin^2(\beta_i) = 1-\sum_{i=1}^{m}p_i \cos^2(\beta_i),
\end{align*}
 so in order to minimize $\Omega_1$, equivalently, we can maximize the following
 \begin{align*}
 \inf_{\yv\in \mathbb{R}^{n},\yv\ne \mathbf{0}}\sum_{i=1}^{m}p_i \cos^2(\beta_i).
 \end{align*}
If we restrict $\|\yv\|_2 = 1$, then we have that
 \begin{align*}
 \cos^2(\beta_i) = \yv^{T} \bv_i\bv_i^{T}\yv.
 \end{align*}
 Therefore
  \begin{align*}
 \sum_{i=1}^{m}p_i \cos^2(\beta_i) = \sum_{i=1}^{m}p_i\yv^{T} \bv_i\bv_i^{T}\yv,
 \end{align*}
 where the right hand side equals
   \begin{align*}
\yv^{T} B^{T}\diag(\pv) B\yv.
 \end{align*}
 Notice that
 \begin{align*}
 \min_{\yv \in \mathbb{R}^{n}, \|\yv\|_2 = 1}\yv^{T} B^{T}\diag(\pv) B\yv = \sigma_n(B^{T}\diag(\pv) B),
 \end{align*}
 in which $\sigma_n(\cdot)$ denotes the smallest singular value of the matrix. The previous discussions can be summarized as:
 \begin{Theorem}
    \begin{align}
 \Omega_1 = 1 - \sigma_n(B^{T}\diag(\pv) B).
  \end{align}

 \end{Theorem}

 Similarly, we have that:
 \begin{Corollary}
    \begin{align}
 \Omega_2 = 1 - \sigma_1(B^{T}\diag(\pv) B),
  \end{align}
  in which $\sigma_1(\cdot)$ denotes the maximal singular value of the matrix.
 \end{Corollary}

 Notice that minimizing $\Omega_1$ is equivalent to maximizing $\sigma_n(B^{T}\diag(\pv) B)$, then we can solve the following problem instead:
\begin{alignat}{2}
\label{opt.max}
 &\max_{\pv \in \mathbb{R}^{m}} &&\ \sigma_n(B^{T}\diag(\pv) B) \\ \nonumber
 & s.t. &&\ \mathbf{1}^{T} \pv = 1; \\ \nonumber
 &      &&\  p_i \ge 0, \ i = 1, \ldots, m. \nonumber
\end{alignat}

This problem can be rewritten as the following SDP problem, in which $\hat{t}$ denotes the optimized $\sigma_n$ and $\hat{\pv}$ denotes the corresponding probability distribution vector:
 \begin{alignat}{2}
 \label{eq.opt1}
 &(\hat{\pv}, \hat{t})  = &&\argmax_{\pv \in \mathbb{R}^{m}, t \in \RM}\  t  \\  \nonumber
 &s.t.                    && \mathbf{1}^{T} \pv = 1; \\ \nonumber
 &                        && p_i \ge 0, \ i = 1, \ldots, m; \\ \nonumber
 &                        &&  B^{T}\diag(\pv) B - tI_n \succeq 0.  \nonumber
\end{alignat}

After solving the optimization problem of (\ref{eq.opt1}), $\hat{\pv}$ is applied to the RKA to select the rows. Such a scheme will be abbreviated as ORKA in the following.
 \begin{Remark}
 There exist cases such that $\Omega_1 = \Omega_2$, i.e. there exists a vector $\pv$, such that $$\sigma_1(B^{T}\diag(\pv) B) = \sigma_n(B^{T}\diag(\pv) B),$$ i.e. $B^{T}\diag(\pv) B = \frac{1}{n} I_n.$  In such cases, $\Omega_1 = \Omega_2 =1- \frac{1}{n}$, and the optimized probability distribution obtained by solving eq. (\ref{eq.opt1}) is the same as suggested in \cite{c1}. It can be verified that when the columns of $A$ are orthogonal and of equal norm, then such property will hold. $\blacksquare$
 \end{Remark}

\begin{Remark}
The optimization problem (\ref{eq.opt1}) can also be formulated as
 \begin{alignat}{2}
  \label{eq.opt2}
 &\hat{\qv} = &&\ \argmin_{\qv \in \mathbb{R}^{m}}\  \mathbf{1}^{T}\qv  \\ \nonumber
 & s.t.         &&\ B^{T}\diag(\qv) B - I_n \succeq 0; \\ \nonumber
 &              && \ q_i \ge 0, \ i = 1, \ldots, m. \nonumber
\end{alignat}
in the sense that $\hat{t} = \frac{1}{\mathbf{1}^{T} \hat{\qv}}$ and $ \hat{\pv} = \hat{t} \hat{\qv}.$

Since $\qv$ in (\ref{eq.opt2}) is nonnegative, one has that $\mathbf{1}^{T}\qv = \|\qv\|_1$. It is known that the $l_1$ norm minimization problem is likely to return sparse solutions\cite{c6}, which gives that $\hat{\qv}$ is likely to be sparse. In the experiment section, we will also illustrate this phenomena. $\blacksquare$
\end{Remark}

Next, we discuss the relation between the ORKA and the RKA. It is obvious that the projection operations in (\ref{eq.rec}) depend only on the corresponding normal vectors of the hyperplanes $\{H_i\}_{i=1}^{m}$, so we can optimize $\kappa(A)= \|A\|_F\|A^{\dagger}\|_2$ subject to the norms of the rows of matrix $A$. The optimization problem is given as
 \begin{alignat}{2}
 \min_{\{\|\av_i\|_2\}_{i=1}^{m} } \ \   \kappa(A)= \|A\|_F\|A^{\dagger}\|_2.  \nonumber
\end{alignat}

Define $\qv \in \mathbb{R}^m$, in which $q_i = \|\av_i\|_2^2$ for $i = 1\cdots m.$ Then the previous optimization problem can be written as
  \begin{alignat}{2}
 \min_{\qv } \ \   \frac{ \sqrt{\mathbf{1}^{T} \qv}}{\sigma_n(A)}.  \nonumber
\end{alignat}

Set $\mathbf{1}^{T} \qv = 1$ and notice the fact that $A^{T}A = B^{T}\diag(\qv) B $, then we can rewrite the previous problem as follows
 \begin{alignat}{2}
 \label{eq.opt3}
 &(\hat{\qv}, \hat{\sigma}_n)  = &&\argmax_{\qv \in \mathbb{R}^{m}, \sigma_n(A) \in \RM}\  \sigma_n^2(A)  \\  \nonumber
 &s.t.                    && \mathbf{1}^{T} \qv = 1; \\ \nonumber
 &                        && q_i \ge 0, \ i = 1, \ldots, m; \\ \nonumber
 &                        &&  B^{T}\diag(\qv) B - \sigma_n^2(A)I_n \succeq 0.  \nonumber
\end{alignat}
It can be observed that this optimization is equivalent to  the problem given by (\ref{eq.opt1}).

We conclude this observation in the following theorem.
\begin{Theorem}
The ORKA can do at least as good as the RKA, in the sense that if we optimize $\kappa(A)$ over the norms of rows of $A$, we obtain the same probability distribution vector as the one obtained by the ORKA.
\end{Theorem}
\end{section}

\begin{section}{Further Discussions}
Note that although the formulation in (\ref{eq.opt1}) is convex, it is still time consuming to solve this SDP optimization problem. In this section, we will discuss two possibilities to solve it approximately , which can alleviate some of the computational cost. One approximation of (\ref{eq.opt1}) is obtained by relaxing the constraint $B^{T}\diag(\pv) B - tI_n \succeq 0$ by the following linear constraints:
\begin{align}
\label{lpopt}
\bv_i^{T}\diag(\pv)\bv_i \ge t; \forall i = 1, \ldots, m.
\end{align}
It is due to the fact that, for two positive semidefinite matrices $P_1, P_2 \in \mathbb{R}^{n\times n}$, if $P_1\succeq P_2 $, then $P_1(i,i) \ge P_2(i,i)$ holds for $i$ = $1, \cdots, n$. Such relaxation reduces the SDP problem into a Linear Programming (LP) problem, which is computationally easier to solve.

In order to get a better relaxation, we introduce another approximation method which relates to the research of {\em Optimal Input Design} \cite{c10}. Notice that $tr (B^{T}\diag(\pv)B) = 1$, i.e. the summation of all the singular values of $B^{T}\diag(\pv)B$ is fixed, then maximizing $\sigma_n(B^{T}\diag(\pv)B)$ means that we want all the singular values of $B^{T}\diag(\pv)B$ to be close. This leads us to consider maximizing the product of the singular values of  $B^{T}\diag(\pv)B$, or maximizing the determinant of $B^{T}\diag(\pv)B$. As the $\log$ function is monotonically increasing, we can optimize the following
\begin{align}
\max_{\pv \in \mathbb{R}^{m}} \log|B^{T}\diag(\pv)B|,
\label{eq.dopt}
\end{align}
in which $|\cdot|$ denotes the matrix determinant. Optimizing this quantity subject to the same constraints of (\ref{opt.max}) boils down to solve the so-called  {\em D-Optimal Design} problem. One simple iterative algorithm to solve such problem has been suggested in \cite{c17}, which is given as
\begin{align}
p_i^{0} = \frac{\|\av_i\|^2}{\|A\|_F^2}; \ i = 1, \ldots, m; \nonumber \\
p_i^{t+1} = p_i^{t} \frac{\bv_i^{T}(B^{T}\diag(\pv^{t})B)^{-1}\bv_i}{n};  \ i = 1, \ldots, m.
\label{eq.itr}
\end{align}

Here, $\pv^{t}$ denotes the estimation at time $t$, and $p_i^{t}$ denotes its $i$-th element. It has been proven in \cite{c20} that for this algorithm, $\log|B^{T}\diag(\pv^{t})B|$ decreases monotonically w.r.t. $t$. We will make use of such property to approximately solve (\ref{opt.max}) when the objective function is replace by (\ref{eq.dopt}). More discussions will be given in next section.

\end{section}

\begin{section}{Experiments}
In this section, we will conduct experiments to illustrate the efficacy of the presented methods. The setup of our experiment is given as follows. The matrix $A$ is first generated by {\em randn(m,n)} in Matlab with $m=200$ and $n = 20$, after that, each row is normalized, and then scaled with a random number which is uniformly distributed in $[0,1]$. The reason for generating $A$ as such is that in the first stage, the generated rows of $A$ will have different directions which are uniformly distributed on the sphere $S^{n-1}$\cite{c16}; and in the second stage, different rows of $A$ with be assigned with different norms, which is directly related to the probability distribution vector chosen in \cite{c1}.  $\xv$ is generated by {\em randn(n,1)}, and $\bv$ is generated as $\bv = A\xv$. We will compare the Mean Square Error (MSE) along the projection path obtained by all these methods, the first is the one suggested in \cite{c1}
(abbreviated as {\em RKA}), the second is the one obtained by the SDP optimization given by (\ref{eq.opt1}) (abbreviated as {\em ORKA}) and the third is the one obtained by the LP approximations given by (\ref{lpopt}) (abbreviated as {\em LPORKA}), the last is the one obtained by the iterative method to solve the D-Optimal Design criteria (abbreviated as {\em ITEORKA}). We iterate (\ref{eq.itr}) for $10$ times in this experiment. For each method, we run the experiment 2000 times to get the averaged performance.  The CVX toolbox\footnote[1]{http://cvxr.com/} is used to solve the SDP and LP optimization problems. From the experiment, we can observe that the time for solving the LP problem in LPORKA is close to the time needed for the $10$ iterations of (\ref{eq.itr}), and the time needed for solving (\ref{eq.opt1}) in ORKA is approximately 7 times as them.

\begin{figure}[htbp] %  figure placement: here, top, bottom, or page
   \centering
   \includegraphics[width=4.5in]{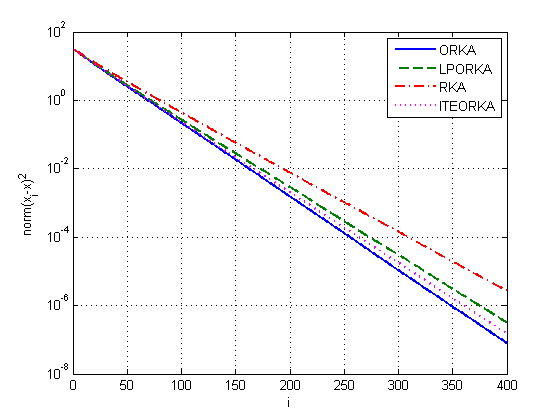}
   \caption{The curves demonstrate the MSE for different methods.   We can see that the ORKA improves the convergence speed the most; the LPORKA method and the ITERKA method also improve the convergence speed, and the ITEORKA method improves more than the LPORKA method. }
   \label{fig.mse}
\end{figure}
\begin{figure}[htbp] %  figure placement: here, top, bottom, or page
   \centering
   \includegraphics[width=4.5in]{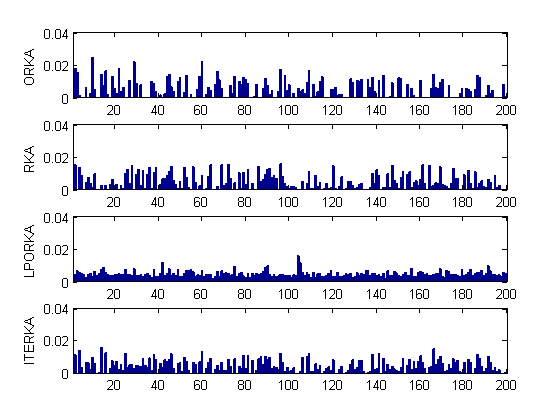}
   \caption{An illustration of the probability distribution vectors obtained by different methods. Note that there are 68 zero elements of the probability distribution vector obtained by the ORKA method, which is $34\%$ sparsity of the total length.}
   \label{fig.rowsel}
\end{figure}

%Then, we will demonstrate a simple example to verify the findings given Remark 2. The matrix $A$ is given by a partial matrix of one orthonormal matrix, and the dimension is $200\times 20$. As shown by the results in Remark 2, the convergence rate of the ORKA will be $1-\frac{1}{20} = 0.95$. We run the projection up to $50$ steps and average out 2000 independent simulations to get an estimation of MSE of the solution at each step. The experiment result is given in Fig. (\ref{fig.untf}).
%\begin{figure}[htbp] %  figure placement: here, top, bottom, or page
%   \centering
%   \includegraphics[width=3in]{test_untf.png}
%   \caption{This figure illustrates the results given in Remark 2. As can be seen, the convergence rate is given by $exp(-0.050736)$, which is approximately 0.0905, this backs up our finding there.}
%   \label{fig.untf}
%\end{figure}
\end{section}
\begin{section}{Conclusion}
This note discusses the possibility and methodology to find a probability distribution vector for selecting the rows of $A$ to result in a better convergence speed of the Randomize Kaczmarz Algorithm. The lower bound and upper bound for the convergence speed is derived first. Then an optimized probability distribution vector is obtained by minimizing the upper bound, which turns to be given by solving a convex optimization problem. Properties of the approach are also discussed along the note.
\end{section}

\end{document}